# Nanostructural polymorphism in the low-birefringence chiral phase of an achiral bent-shaped dimer

*Khoa V. Le,* [1,2,*] *Michael R. Tuchband,* [3] *Hiroshi Iwayama,* [4] *Yoichi Takanishi,* [5] *Noel A. Clark,* [3] and *Fumito Araoka* [1,*]

[1] *RIKEN Center for Emergent Matter Science (CEMS), 2-1 Hirosawa, Wako, Saitama 351-0198, Japan*
[2] *Department of Chemistry, Faculty of Science, Tokyo University of Science, 1-3 Kagurazaka, Shinjuku-ku, Tokyo 162-8601, Japan*
[3] *Department of Physics and Soft Materials Research Center, University of Colorado, Boulder, Colorado 80309-0390, USA*
[4] *UVSOR Synchrotron Facility, Institute for Molecular Science, Okazaki 444-8585, Japan*
[5] *Department of Physics, Kyoto University, Kitashirakawaoiwake-cho, Sakyo-ku, Kyoto, 606-8502, Japan.*

Correspondence and requests for materials should be addressed to F.A. (email: fumito.araoka@riken.jp) or to K.V.L. (email: khoa@rs.tus.ac.jp).

**Abstract**
Polymorphism, the phenomenon that a species can exist in many discrete forms, is common in nature, such as hair colors in an animal species, flower colors in a tree species, and blood types in humans, etc. In materials science, it refers to a solid that can exist in multiple forms with different crystalline structures. In the liquid crystals field, however, polymorphism is hard to find because a discontinuous structural variation is basically impossible because of their fluid or partially fluid nature. Herein we show that the B4 and DC phases that for many years have been classified as distinctive phases are connected, in terms of their nano-architectures, based on the study of a single compound, a flexible bent-shaped dimer. The surrounding solvent is the key to assisting the dimeric molecules in morphing and adopting different supramolecular structures at the mesoscale. Furthermore, we accidentally find a novel nanotube-like structure that has not yet been reported in view of the B4/DC phases. Together with the known sponge (DC) and the helical filament (B4) structures, they are just some of the manifestations of the polymorphism in a class of low-birefringence, chiral phase from achiral liquid crystals.

The processes underlying the self-assembly of constituent building blocks into complex, hierarchical superstructures remain among the most intriguing mysteries in materials science today. This research field has been actively pursued and widely exploited to develop novel functional materials, including both nature-inspired biomimetic materials and metamaterials [1] [2]. Controlling the shape, size, orientation, and arrangement of constituents at the molecular level, giving rise to desirable emergent properties at macroscopic scales is, however, a significant challenge. Organization can arise spontaneously through a variety of non-covalent intermolecular forces [3] or can be induced by external factors such as electric and magnetic fields, temperature, light, pH, or interactions with foreign molecules through the so-called solvent or dopant effects, examples of which are aqueous systems of peptide- or glycolipid amphiphiles [4] [5] [6]. At first glance, one might imagine that the constituent building blocks of a material made up of chiral superstructures must be chiral, as a manifestation of chiral amplification. In this context, the self-assembly of achiral bent-core molecules is unique and interesting, since they can geometrically pack to form chiral and polar symmetry-breaking phases with large conglomerate domains over micron dimensions or larger [7] [8] [9] [10]. Hough et al. reported that in some bent-core molecules which have intralayer semi-crystalline order, a twisted helical nanofilament (HNF), called the B4 phase, or lamellar sponge (SP) structure, called the dark conglomerate (DC) phase, can form [11] [12]. The B4 phase is particularly intriguing because it has strong optical activity [13] [14], large effective nonlinear optical coefficients [15] [16], strong gelation ability [17] [18], enhanced hydrophobicity [19], anisotropic charge transport properties in photovoltaic devices [20], and structural colors [21] [22] [23]. One can envision a wide range of applications for this material, such as tunable optical rotators, piezoelectric elements, chiral detectors, asymmetric chemical syntheses, ultra-dry surfaces, solar cells, color reflectors, etc., among other possibilities [10] [24] [25] [26]. The formation of the dark conglomerate SP and HNF structures appear to be driven by the same mechanism, i.e., a mismatch of the two half-layers formed by the molecular arms of the compound in a smectic layer induces saddle-splay curvature to relieve the elastic strain caused by their mismatch. This leads to the spontaneous formations of left- and right-handed HNFs or dark conglomerate SP domains at the macroscopic scale with a huge optical activity, on the order of 1 deg/μm. However, this still leaves open the following question: what causes a particular material to adopt either the SP or the HNF structure?

In this work, and for the first time to our knowledge, we report that an achiral bent dimer molecule which exhibits an SP morphology as the ground state structure can be continuously transformed to an HNF structure and an unusual hollow nanotube (tubular) structure simply by making mixtures with either mesogenic or organic isotropic solvents and tuning the concentration. We characterize this behavior with polarizing optical microscopy (POM), scanning electron microscopy (SEM), freeze-fracture transmission electron microscopy (FFTEM), atomic force microscopy (AFM) and synchrotron X-ray diffraction (XRD). The thorough understanding of the mechanism of this polymorphism may provide not only insight into the nature of the formation of the B4 "banana" phase and other bent-core liquid crystal phases but may also shed light on the materials science pursuit of bottom-up supramolecular design approaches which exhibit multiple tiers of self-assembly.

**Results**

**Neat material 12OAz5AzO12.** In this work, we study a bent-shaped symmetric dimer compound which we call 12OAz5AzO12. It has a pentamethylene spacer connecting two mesogenic wings, namely, 4-n-alkoxyazobenzene-4'-carbonyloxy-n-dodecanes (Fig. 1a) [27]. The odd number of methylene units in the linkage makes the molecule adopt a bent shape in the all-trans conformation. Its phase sequence is Iso (108°C) SmC$_A$ (94.1°C) SmX, where the SmX phase is a low birefringence phase with chiral conglomerate domains and is stable down to room temperature. This compound has been utilized in earlier works to investigate chirality control in the DC/B4 phase [28], twisted nematic director fields [29], and biomolecular adsorbates [30]. Upon cooling to the SmX phase (Fig. 1b,c), POM reveals a very dark texture with some birefringent inclusions which are most probably thin layers of SmC$_A$-like structures pinned on the surfaces remaining from the higher temperature SmC$_A$ phase, as their retardation does not change with cell thickness. By slightly decrossing the polarizers, the spontaneous chiral resolution can be easily checked (Fig. 1d,e). When the SmX phase forms, cracks in the material appear due to volume shrinkage, which is common in the B4 phase (Fig. 1d,e). For many years, this SmX phase was identified as the 'B4 phase' since it is a solid, low-birefringence phase that appears below a smectic phase and shows no response to electric field, all characteristics of a B4 phase. However, for the first time, our SEM and FFTEM observations show that neat 12OAz5AzO12 has a sponge structure, rather closer in morphology to the DC phase (Fig. 3a,e). We observe only disordered focal conic domains with clear saddle-splay curvature throughout the bulk sample. In particular, when the sample is fractured near the glass interface, the layers stand on the surface with the layer normals parallel to the glass surface [Fig. S1], as reported in ref. [12]. These observations correspond well with those reported by Hough et al. for a DC phase nanostructure [12]. Observations using AFM also support the suggestion of a sponge structure [Fig. S2].

**Mixtures of 12OAz5AzO12 and a rod-like nematic liquid crystal (NLC) ZLI-2293.** We made mixtures of different concentrations of 12OAz5AzO12 with a rod-like nematic material called ZLI-2293 (Merck) to observe the various effects on the nanoscale morphology of the phase. ZLI-2293 was chosen because it has a wide nematic temperature range (91°C to below room temperature). The phase diagram of the mixtures in Fig 2a is based on POM observations. The $SmC_A$ phase is completely suppressed at 20 wt% ZLI-2293, with this same behavior reported in other systems of bent-core LCs with rod-like LC solvents [31] [32] [33]. At higher concentrations of ZLI-2293 (>20 wt%), all the mixtures exhibit two distinct phase transitions: the Iso–N transition and a transition from the homogenous N phase to the composite SmX phase, composed of nano/micro-phase segregated 12OAz5AzO12 and nematic ZLI-2293 [14] [32]. Fig. 2b contains a POM image of a contact cell at room temperature demonstrating continuous changes in texture along the concentration gradient of ZLI-2293. On the right-hand side of the image, the neat 12OAz5AzO12 exhibits its characteristic dark texture with some birefringent inclusions (region A) similar to those in Fig. 1c. On increasing the ZLI-2293 concentration, the texture becomes darker and smoother with no more birefringent inclusions (region B). Further increasing the ZLI-2293 concentration, we observe many dendrites growing and separating from the nematic background (region C). Finally, in region D, the ZLI-2293-rich nematic phase exhibits a Schlieren texture. Decrossing the polarizers reveals that the conglomerate domains in region A persist into region B and even C [see more in Fig. S3]. But how could such a sponge structure nucleate, develop, and phase separate from the nematic background? To our surprise, SEM and FFTEM observations reveal that the nano-phase segregated 12OAz5AzO12 in the mixtures (observed after ZLI-2293 is removed, see Methods) no longer has the SP structure, but transforms into an HNF morphology with increasing concentration of ZLI-2293 (up to ~60 wt%) (Fig. 3c). Even more surprisingly, at a very high concentration of ZLI-2293 (≈ 80 wt% or above) a hollow nanotube structure of diameter ≈ 100 nm, slightly larger than that of the HNFs (70 – 80 nm), predominantly forms (Fig. 3d) [see also Fig. S4].

As the SEM resolution is generally quite limited for organic materials, FFTEM observations were also performed on these mixtures to visualize the layering, which should be approximately 4 – 5 nm. For the mixture with 60 wt% ZLI-2293, in which the HNFs are dominant, the HNF layer stacks bending and twisting with saddle-splay curvature is clearly observed (Fig. 3g). We also observe several HNFs branching out from a single HNF, a process that enables the chirality preserving growth which leads to the macroscale chiral conglomerate domains in the B4 phase [34]. We also observe several examples of coherently twisting HNFs, with the layers in adjacent filaments face-to-face [Fig. S5]. A larger scale view of the HNF morphology [Fig. S5] demonstrates that the filaments can grow independently or collectively in the bulk and form a nanoporous structure, as expected for HNFs [11]. These structures are undoubtedly HNFs, but are remarkably different from those formed by traditional bent-core molecules [11] [34] [35], as analyzed from FFTEM data: (i) the cross-section of the HNFs formed of 12OAz5AzO12 is rectangular with the dimension along the layer normal (≈ 80 nm) about 4 times longer than the width (≈ 20 nm), as opposed to a nearly square cross-section in other bent-core LCs, and (ii) the layers (each layer ≈ 5 nm) stack such that a layer 'shelf' is evident every 3 – 4 layers, with the layer shelves stack on one another while slightly shifting towards to the edges of the filament (the direction normal to the filament axis). On the other hand, when the concentration of ZLI-2293 is between 20 – 40 wt%, neither a clear HNF or SP structure is discernable; rather, an intermediate morphology between SP and HNF observed (Fig. 3b,f). The presence of this transitional regime demonstrates that the composite SmX phase of chiral conglomerate domains is a spectrum of morphologies that exist between the SP and HNF structures. All these morphologies have a saddle-splay curvature (negative Gaussian curvature) and are identical from a topological viewpoint, i.e., catenoid and helicoids (see Supporting Information of ref. [35]), respectively, and blends of the two in the intermediate regime. This shows that the interfacial tension between these two immiscible components plays an essential role in determining the morphology. Finally, in good accordance with SEM, when the concentration of ZLI-2293 is high (80 wt% or above), almost solely nanotubes are observed (Fig 3h). Their surfaces are very smooth smectic layers with *zero* Gaussian curvature, as also confirmed by AFM [Fig. S6]. Some of the nanotubes are even connected with the twisted HNF structures, indicating that the mechanisms of the formation of the nanotubes and HNFs are closely related. Here also, we believe it is the interfacial tension that transforms the HNF into the nanotube structure.

Our experiments with other mesogenic solvents (5CB, CB15) and non-mesogenic solvents (n-dodecane, chloroform) also show such polymorphic behavior, with lower concentrations of the non-mesogenic guest material necessary to form the nanotube structures [Fig. S7]. In particular, we also find that nanotube structures even form at extremely low concentrations of 12OAz5AzO12, e.g., 98.7 wt% n-dodecane [Fig. S8]. We stress that if the starting material has the HNF structure, no matter how much solvents are mixed, the HNF structure always remains, either in the form of helical twists or helical ribbons [17] [18].

**X-ray diffraction and resonant carbon K-edge X-ray scattering.**
To determine the nature of the molecular packing within these various nanostructures, we performed synchrotron X-ray diffraction experiments. Fig. 4a shows the data taken on three samples with the distinct morphologies after the solvent was removed. Curiously, they show essentially the same behavior, except for some variations in the peak widths of particular features. At least 8 different harmonics of the first-order diffraction attributed to the layer periodicity ($\approx$ 5 nm) appear at small angles, signifying the robust lamellar ordering. The full width at half maximum of these peaks reveals the finite size of the lamellar correlations to be is about 20 layers [11] [18], which agrees well with the results obtained from the electron microscopy observations. At wide angles, the numerous peaks point to long-range crystalline or semi-crystalline order within the layers (Fig. 3) and that 12OAz5AzO12 maintains this rigid ordering when nano-/micro-phase segregating from the guest solvent, regardless of the concentration.

The resonant soft X-ray scattering (RSoXS) at the carbon K-edge (284.5 eV), which is a powerful technique to probe the periodic modulation of the orientation of carbon bonds [36] [37], shows a clear distinction between the three morphologies (Fig. 4b). HNFs, appearing in the middle doping range, show relatively sharp scattering peaks at the scattering angle corresponding to the half-pitch ($\approx$150 nm) periodicity of the helical filaments, which is consistent with that observed by FFTEM (Fig. 3g) and analogous to that in another dimer with similar structure [26]. On the other hand, the SP structures show broad peaks at the slightly smaller scattering angle region, which could be corresponding to the average distances between the lamellar inter-connections. Interestingly, the nanotubes show just scarce signals in this region, meaning almost no nanoscopic structural periodicity.

**Discussion**

A study by Lin et al. showed that by mixing a bent-core molecule with a poor solvent (THF/water solution) they could force their assembly into different superstructural organizations, including flat, elongated lamellar crystals, helical ribbons, and tubules [38]. They reported that the helical morphology is driven by conformational chirality of the bent-core molecules which results in the twisting and bending of lamellar crystals into the various complex shapes [39] [40]. More recently, Cano et al. showed that some ionic bent-core dendrimer molecules can aggregate in water to form a variety of assemblies, including rods, spheres, fibers, helical ribbons, or tubules [41]. In our case, mesogenic and non-mesogenic organic guest solvents are used, and 12OAz5AzO12 always completely phase separates from the guest due to the strong crystalline ordering it possesses, as gleaned from the X-ray data (Fig. 4).

The unusual phase sequence that we find in 12OAz5AzO12 (SmC$_A$–DC) has also been observed recently by Chen et al., where a bent-core compound (W624 in ref. [42]) was reported to exhibit a stable DC phase below the B2 phase. This is intriguing because, according to other previous reports, while the B4 phase is rather solid (semi-crystalline) and normally appears below the B2 and/or the B3 phases, the DC phase is more liquid-like and appears below an isotropic liquid and not the ordered SmC$_A$ or B2 phase [12] [11]. The DC phase can in principle be transformed into the ordered B2 phase on application of an electric field [12] [43] or could be induced from the columnar B1$_{rev}$ phase [44]. Like in W624, we observe re-crystallization [Fig. S9] when the DC phase is heated back up to the SmC$_A$ phase. Therefore, the DC phase reported for 12OAz5AzO12 here can be considered to be in a meta-stable state. As such, interfacial tension from the guest solvent can drive it to the HNF structure which would make it the more stable configuration. On the other hand, by continuing to increase the solvent content to some extent, the HNF is then transformed to the nanotube structure. Recently, the observation of the nanotube structures similar to ours has also been reported in mixtures of an acute-angle (ca. 45°) bent-core compound with nematogenic or smectogenic additives over a narrow concentration range (Fig. 3h and Fig. S6) [45]. The acute-angle bent-core compound also exhibits re-crystallization on heating, which implies that molecules exhibiting metastable DC/B4 phases may be suitable candidates to express a polymorphic behavior when mixed with a solvent or influenced by other external factors. Because the nanotube structure does not have a negative Gaussian curvature like the SP or the HNF structures, it appears that the interfacial energy is strong enough to suppress saddle-splay structure and completely change the expected morphology. For instance, when we consider the molecules at the outmost layers of each nanotube, whose exposed chemical groups are the liquid-like dodecanoxy groups (-OC$_{12}$H$_{25}$), the system can gain high entropy even if the dimer molecules strongly aggregated because the outer surface is more soluble with the alkylated portions of the solvent molecules [46]. This also explains why we require less n-dodecane (C$_{12}$H$_{26}$) to form the nanotubes.

From the symmetry viewpoint, Fig. 5 shows how a change in local in-plane hexatic lattice symmetry, from a square lattice in the layer mid plane to a rectangular lattice in the layer mid plane, mediates a change in growth morphology from twisted ribbon B4 to cylindrical ribbon nanotube. In the typical B4 twisted ribbon case the upper and lower half layers are structurally identical, related by a 2-fold rotation around **P**. The upper and lower half layers relate identically to the overall ribbon geometry. In the nanotube (cylindrical ribbon) case, the lower symmetry implies a structural difference between the upper and lower half layers, e.g.. a larger tilt in the upper half than that of the lower. In this case the orange lattice must be rectangular and the upper and lower half layers relate differently to the overall structure with one along the ribbon and the other normal to it. In both cases the filament edges are a low Miller index face of the 2D lattice, 11 in the case of the ribbon and 10 in the case of the nanotube.

This study demonstrates the link between the various apparently distinct nanostructures exhibited by 12OAz5AzO12. By making mixtures of this compound with different organic solvents, we can continuously tune the morphology from sponge to HNF to nanotubes, even accessing the states in between which seem to blend the nanostructures of the morphologies between which it resides. This indicates that the DC and B4 phases may not necessarily be distinct from each other or that they may be best described by unifying them into a single phase, though it is beyond the scope of this work to do so.

The ease with which we can tune the nanostructures manifested in the mixtures illustrates the special nature of the 12OAz5AzO12 molecule, which is the only flexible bent molecule that forms the DC *and* HNF phases to our knowledge. All other molecules which form either of these phases are of the rigid bent-core variety. The flexibility of the alkyl linker may permit 12OAz5AzO12 to adopt a wider variety of molecular conformations which stabilize the multiple nanoscale structures which we observe in the mixtures. More detailed molecular-scale experiments and modeling may be required to demonstrate this proposal conclusively.

In summary, we have shown that we can tune the morphology of 12OAz5AzO12 from the sponge to helical nanofilament to nanotube structures simply by introducing an organic solvent guest. The structural transformation is explained in terms of the interfacial tension and entropy-driven processes. The sponge and the helical nanofilament structures are chiral, but the chiral nature of the nanotubes has not yet been confirmed. However, we note that the dark conglomerate chiral domains can persist even into the region of the nanotube structure (Fig. S3), so they may be chiral as well. This work demonstrates the rich and diverse polymorphism exhibited by this achiral bent molecule [47], which may find applications in the context of porous materials, patterned materials, and photonic metamaterials, and contribute to understanding other research topics such as organo-gels, host-guest chemistry, asymmetric chemical syntheses, and chiral recognition and selection.

**Methods**

**Sample preparation and texture observation.** 12OAz5OAz12 was synthesized and supplied by DIC Corporation. The material was mixed with ZLI-2293 at different concentrations (20, 40, 60, 80, 90 wt%) by dissolving them together in chloroform. After the chloroform evaporated, the mixtures were heated up to 120°C in an oven and then cooled slowly down to room temperature. Mixtures with the other studied solvents were prepared with the same protocol. The samples were introduced into commercial glass cells (typical thickness ca. 5 μm) or placed between two glass slides and the optical textures were observed under Nikon polarizing microscopes Eclipse LV100POL or Eclipse E400POL equipped with Mettler Toledo FP90 central processor and an FP82HT hot stage for temperature control.

**SEM, AFM, TEM observations.** For SEM observation, the samples were introduced to glass cells at ~120°C which were fabricated in-house by weakly bonding together two glass plates with a UV-curable glue (Norland NOA81). At room temperature, the glass plates were then detached from each other with the samples remaining on their surfaces. They were then gently placed in n-hexane for several hours to dissolve the guest solvent used for making the mixtures. The glass plates were taken out of the n-hexane and heated at 65°C to evaporate the remaining n-hexane from the sample. The dry samples were then coated with a thin layer of Pt-Pd using a Hitachi ion sputter MC1000 and observed using a Hitachi SEM SU8010. For AFM observation, the samples were prepared under the same conditions as in SEM, except they were not coated with Pt-Pd and they were observed using an Asylum Research Cypher AFM (Atomic Force Microscope). For TEM observation, the mixtures were prepared as described in the sample preparation section and were cast directly onto copper grids. The copper grids were then gently placed into n-hexane to rinse out the guest solvent. After drying, the copper grid was transferred to a JEOL JEM-2100 or JEM-1230 TEM for observations.

**FFTEM observation.** The samples were sandwiched between 2 mm × 3 mm glass planchettes and cooled from the isotropic phase to the desired phase at room temperature. The guest solvent was removed as mentioned above and then quenched to –180°C by sudden immersion in liquid propane. They were then transferred to a freeze-etching system BAL-TEC BAF 060 to be fractured under high vacuum at –140°C. The fractured surfaces were subsequently coated with a 2 nm-thick layer of platinum deposited at an oblique angle (45°), followed by ~20 nm of carbon layer deposited from directly above (90°) to increase the rigidity of the replicas. After dissolving the 12OAz5OAz12 away in methyl acetate, the Pt-C replicas were placed on copper grids and observed in a TEM Philips CM10. Images were taken with a TEM-mounted 1K × 1K Gatan Bioscan digital camera. The surfaces facing the platinum shadowing direction appear darker in the TEM images because they accumulate more platinum. This provides the contrast for visualizing height variations in the surface topography.

**Synchrotron X-ray diffraction.** Synchrotron 2D X-ray diffraction measurements were performed using beamline BL45XU at RIKEN SPring-8 (Hyogo, Japan). After rinsing out the guest solvent with n-hexane, the samples were put into 1.5-mm-diameter glass capillaries and exposed to X-ray at room temperature (20°C) for 50 s. Diffraction data were recorded on an imaging plate (model R-AXIS IV++, Rigaku) detector. The X-ray beam (wavelength 1.00 Å) was mono-chromated with a diamond (1 1 1) double-crystal monochromator. The camera length was 0.40 m. The scattering vector $q$ was calibrated using silver behenate ($d$ = 58.380 Å) as a standard sample.

For soft carbon K-edge X-ray scattering, we used linearly polarized resonant soft X-ray scattering at the carbon K-edge (284.5 eV) at the UVSOR Synchrotron Facility, Institute for Molecular Science (Aichi, Japan). A homemade vacuum chamber with a cooled CCD camera (Newton, Andor) was newly designed for small angle detection ($2\theta \approx$ 1-15˚). The samples (mixtures of 12OAz5OAz12 and ZLI-2293) were drop-casted onto 100-nm thick $Si_3N_4$ membranes (NTT-AT), gently washed by n-hexane to remove ZLI-2293 before exposed to X-ray at room temperature.

**Acknowledgements**
This work was partially supported by JSPS KAKENHI Grant Number JP18K14094 and by NSF MRSEC Grant No. DMR-1420736. We are grateful to Dr. Y. Aoki (DIC Corporation) for kindly supplying the dimer material. We thank Dr. D. Miyajima (Information Transforming Soft Matter Research Unit, RIKEN CEMS) for letting us use SEM. We also thank Dr. N. Horimoto and Dr. Y. Ishida (Emergent Bioinspired Soft Matter Research Team, RIKEN CEMS) for assistance in taking the AFM images and the X-ray diffraction data, and Ms. T. Kikitsu and Mr. D. Inoue (Materials Characterization Support Team, RIKEN CEMS) for assistance in taking TEM and SEM images. We thank Dr. T. Giddings (University of Colorado Boulder) for technical help with FFTEM observations. We also thank other members of Soft Materials Research Center (University of Colorado Boulder) for fruitful discussions. K.V.L. thanks H. Nguyen for helpful advice, Mrs. J. and Mr. P. Acker for their hospitality.

**Author contributions**
K.V.L. performed experiments and analyzed results. F.A. designed and directed the project. M.R.T. performed a portion of the FFTEM observations. F.A., H.I, Y.T. performed the resonant carbon K-edge X-ray scattering. K.V.L., M.R.T., N.A.C., and F.A. explained the results and wrote the manuscript. All authors have seen and approved the final manuscript.

**Additional information**
Supplementary Information accompanies this paper at

Supplementary Figures 1-9

Competing financial interests: The authors declare no competing financial interests.

Reprints and permission information is available online at

How to cite this article:

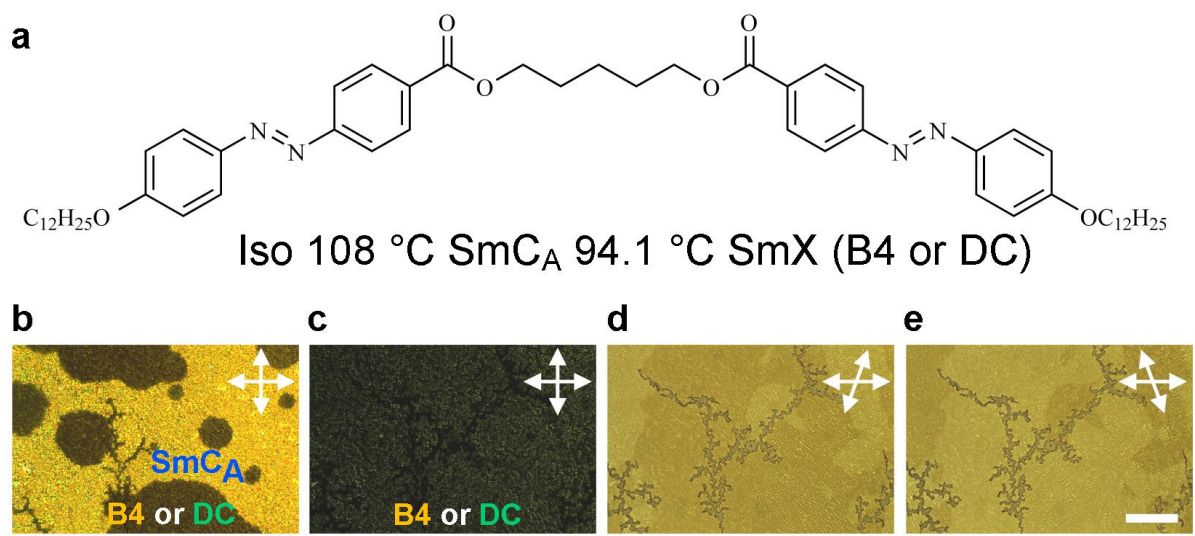

**Figure 1 | Chemical structure, phase sequence, and POM textures of the dimeric compound 12OAz5AzO12.** (**a**) Chemical structure and phase sequence of 12OAz5AzO12. (**b,c**) The transition from the highly birefringent (gold color) SmC$_A$ to the nearly optically isotropic SmX (B4 or DC) phase at 94.1 °C. (**d,e**) Interchanging bright and dark domains upon decrossing the polarizers, indicating a macroscopic chiral resolution of the sample. The cell was composed of two bare glass slides. Scale bar, 200 μm.

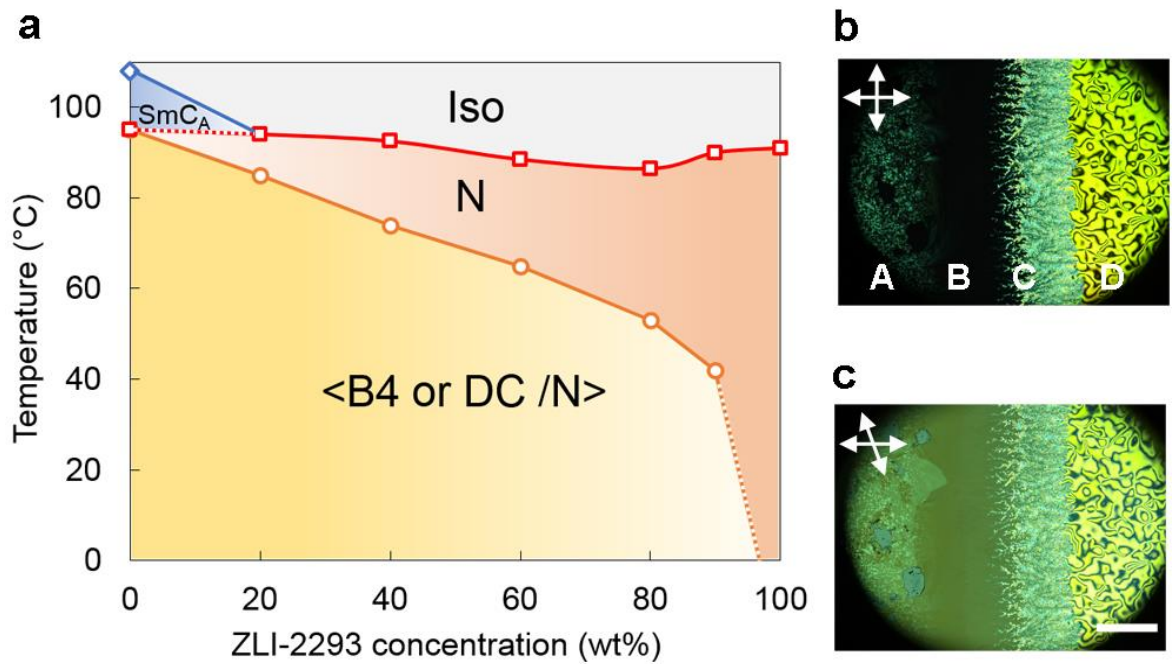

**Figure 2 | Phase diagram of mixtures of 12OAz5AzO12 and rod-like nematic material ZLI-2293 and a contact cell preparation of the two components.** (**a**) Phase diagram of 12OAz5AzO12 and ZLI-2293 in temperature and concentration. (**b,c**) POM textures of a contact cell of 12OAz5AzO12 and ZLI-2293 at 33 °C between crossed and decrossed polarizers. Region A is rich in 12OAz5AzO12 and is low birefringence but with specks of birefringent pinned SmC$_A$ domains, while region D is rich in ZLI-2293 exhibiting the characteristic Schlieren texture of the nematic phase. Region B is much darker and smoother than region A, with no birefringent inclusions. Region C exhibits rather dendritic domain growth, micro-separating from ZLI-2293 that is observable due to the refractive index mismatch with the surrounding ZLI-2293. Scale bar 400 μm.

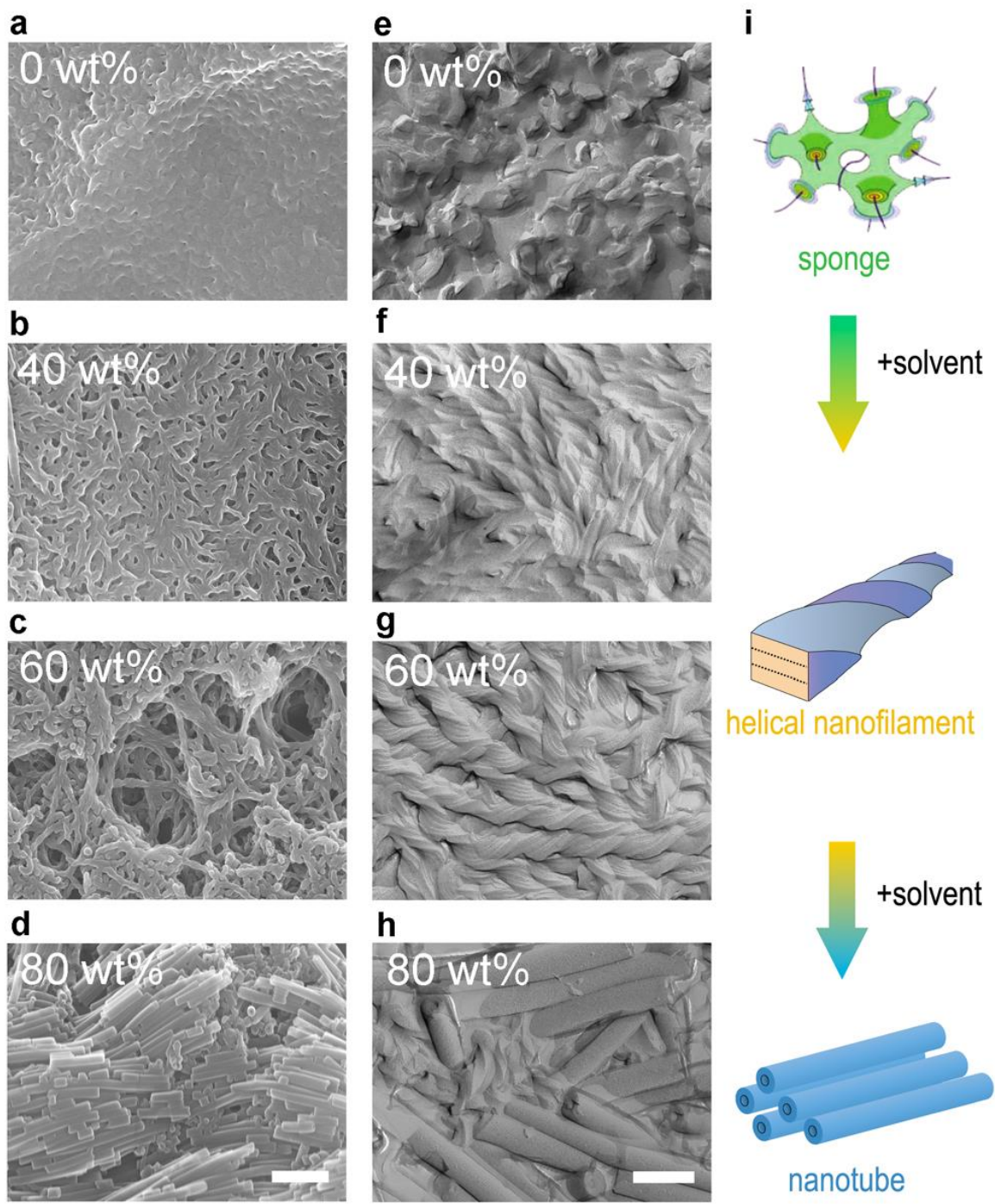

**Figure 3 | SEM and FFTEM images of polymorphism in mixtures of 12OAz5AzO12 and an organic solvent.** (**a-d**) SEM images and (**e-h**) FFTEM images of different nanostructures of the nanophase separated 12OAz5AzO12 formed by mixing with different concentrations of ZLI-2293 at room temperature. Note that the diameter of the nanotube (≈ 100 nm) is slightly larger than that of the HNFs (≈70 – 80 nm). (**i**) Illustrations of the sponge, helical nanofilament, and nanotube morphologies. The sponge structure is adapted from [12], with permission. Scale bar 700 nm in SEM, 200 nm in FFTEM.

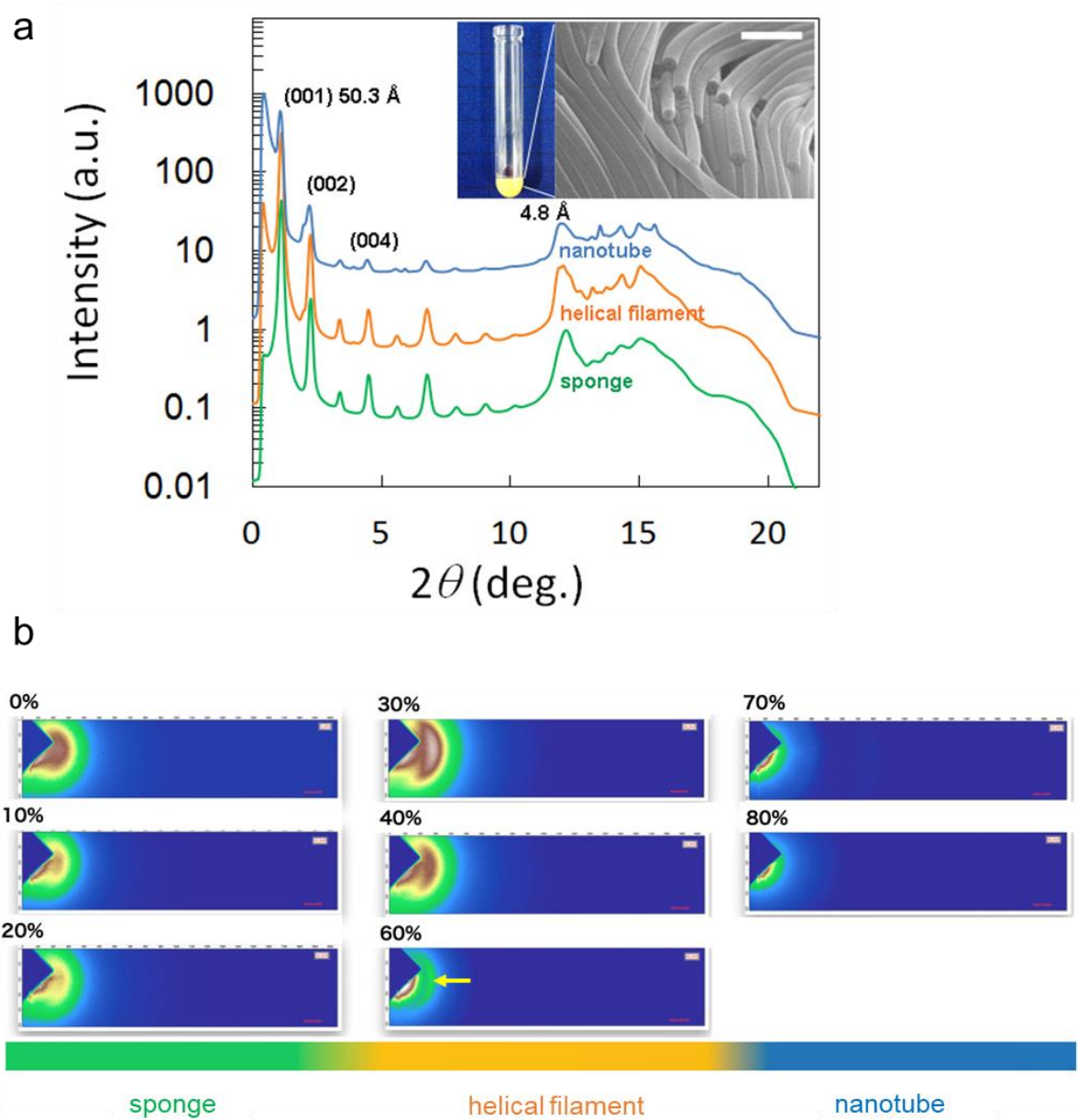

**Figure 4 | X-ray diffraction profiles of three different structures featured by 12OAz5AzO12.** (**a**) The three distinct morphologies exhibit the many orders of reflection in the small angle, corresponding to a high degree of ordering of the layer structure, the spacing of which roughly corresponds to the molecular length of 45.3 Å [27]. Multiple peaks at wide angles indicate long-range crystalline intralayer order. Inset is the xyrogel of 12OAz5AzO12 (after evaporation of the solvent) created from the mixture with ca. 90 wt% n-dodecane. The plots were shifted along the y-axis for ease of comparison. Scale bar in SEM image is 500 nm. (b) RSoXS pattern transformation with respect to the nanoscopic morphological change. Percentage indicates the weight concentration of ZLI-2293 used in the mixtures to create the desired structures. The arrow at 60% marks the peak at $q = 4.2 \times 10^{-3}$ Å$^{-1}$, corresponding to the half-pitch (≈150 nm) of a helical filament. The polarization direction of X-ray beam is vertical.

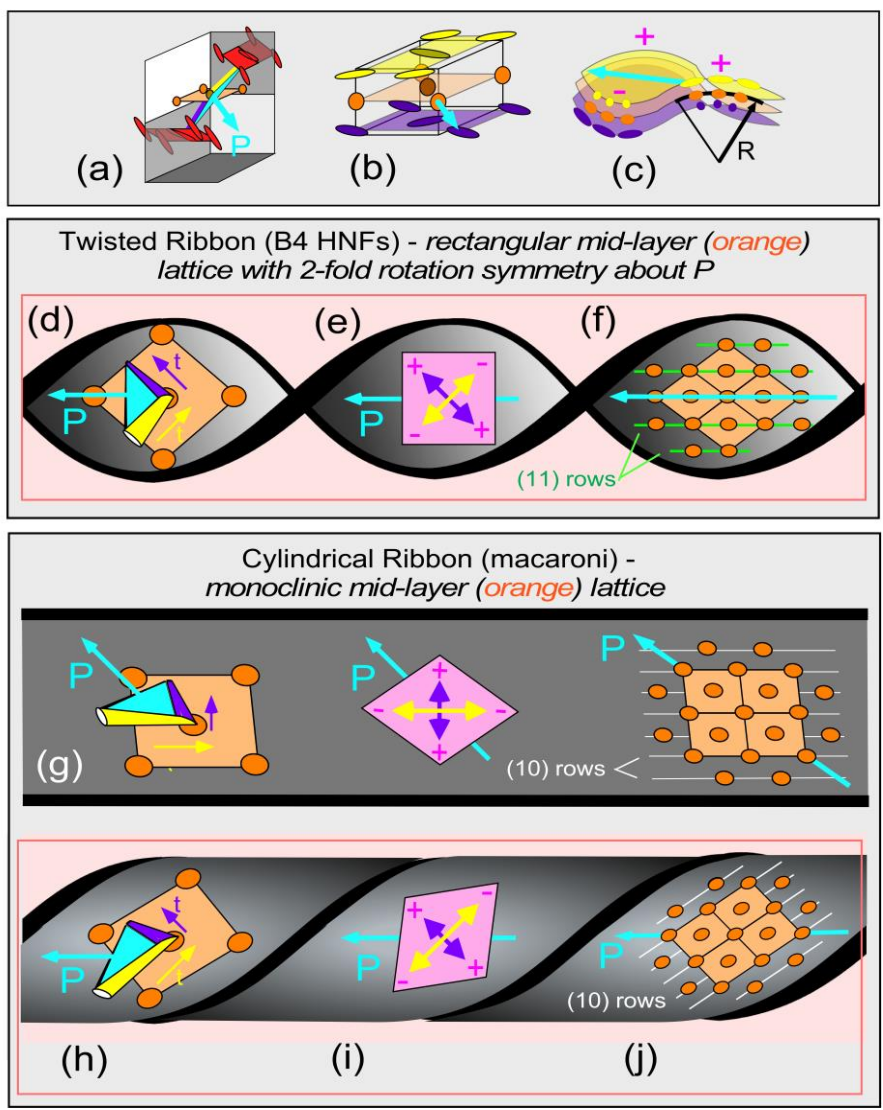

**Figure 5 | Change in growth morphology from twisted ribbon B4 to cylindrical ribbon nanotube.** A change in local in-plane hexatic lattice symmetry, from a square lattice in the layer mid-plane to a rectangular lattice in the layer mid-plane.

SUPPLEMENTARY INFORMATION

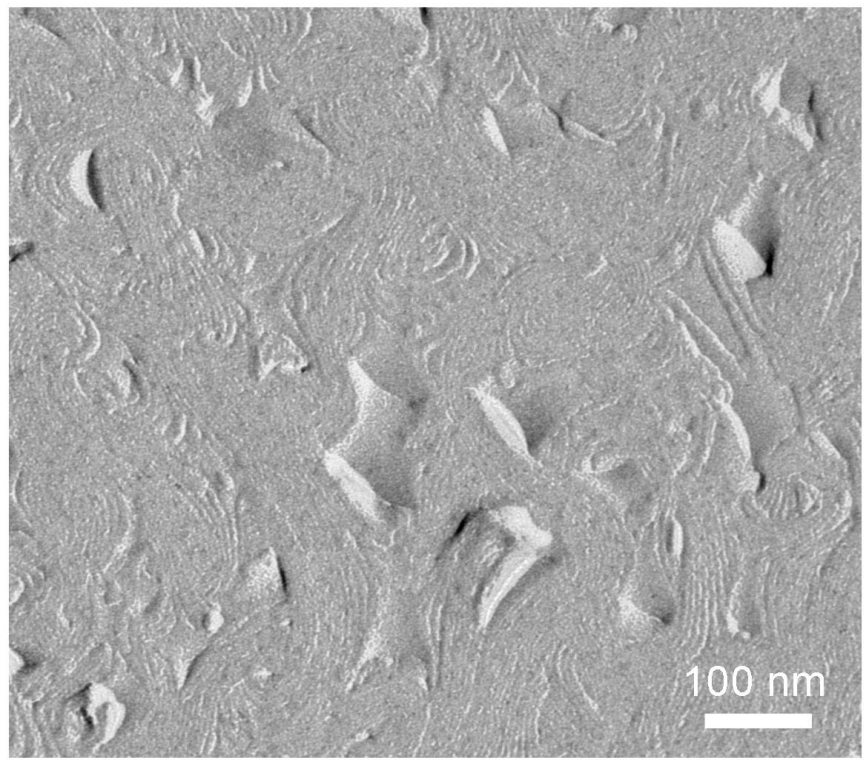

Figure S1. FFTEM image of the neat 12OAz5AzO12 sample where the sample is fractured near a glass surface so that the layers intersecting with the glass surface can be observed as curved lines. Note that even some catenoids that are characteristic of the sponge structure are also observable.

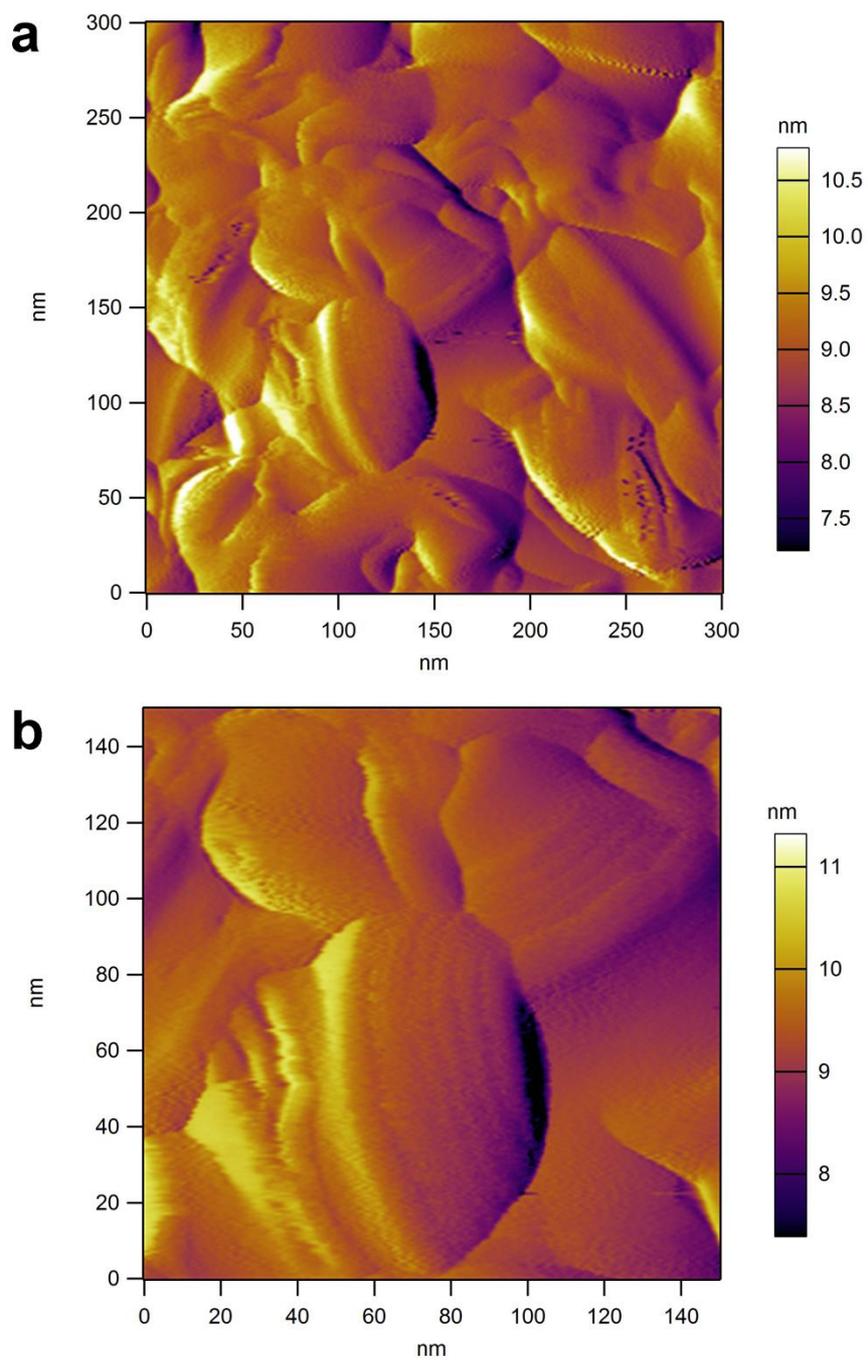

Figure S2. (a) AFM image of the neat 12OAz5AzO12 sample. (b) The center of the image in (a) is enlarged and shown, where the topography with saddle-splay curvature (negative Gaussian curvature) is clearly observed.

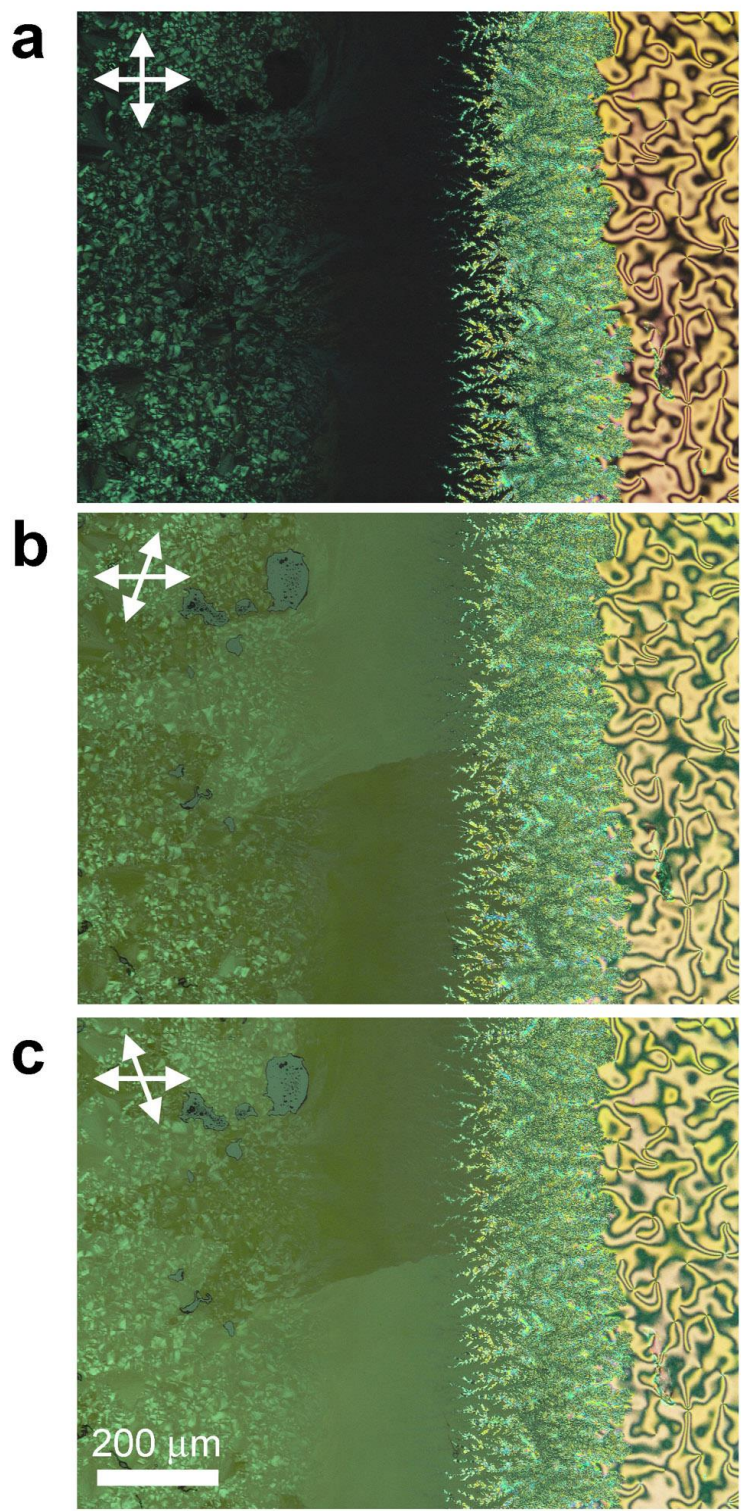

Figure S3. POM textures of a contact cell of 12OAz5AzO12 (left side) and ZLI-2293 (right side) observed at room temperature between (a) crossed and (b,c) decrossed polarizers. Decrossing the polarizers reveals that homochiral conglomerate domains originally from the region of neat 12OAz5AzO12 persist into the region of the mixture with ZLI-2293.

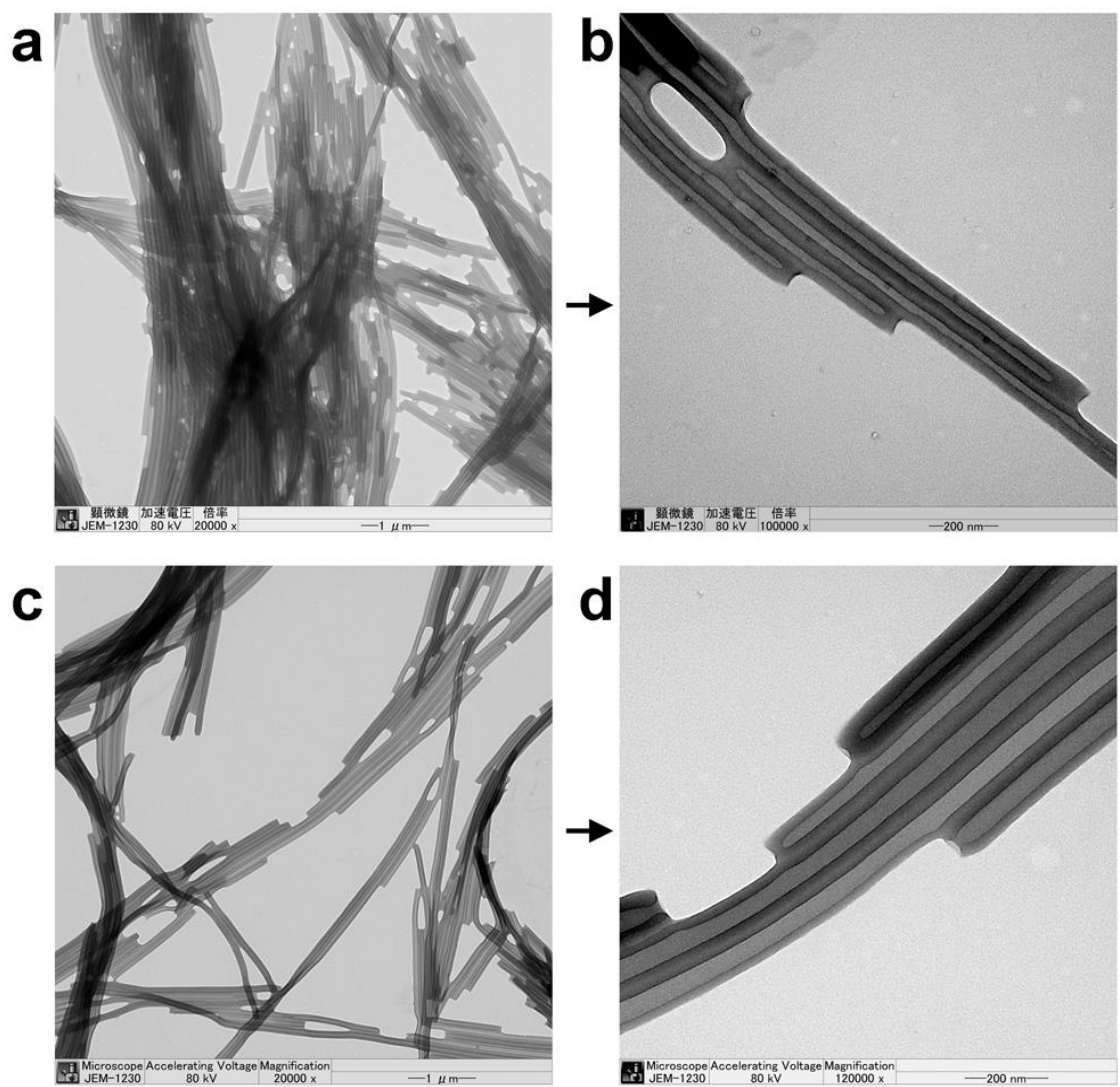

Figure S4. TEM images of hollow nanotubes formed by 12OAz5AzO12 mixed with (a,b) mesogenic solvent ZLI-2293 in a 90 wt% mixture, and with (c,d) non-mesogenic solvent n-dodecane in a 90 wt% mixture. Note that the magnifications in (b) and (d) images are different: 100 000 and 120 000, respectively.

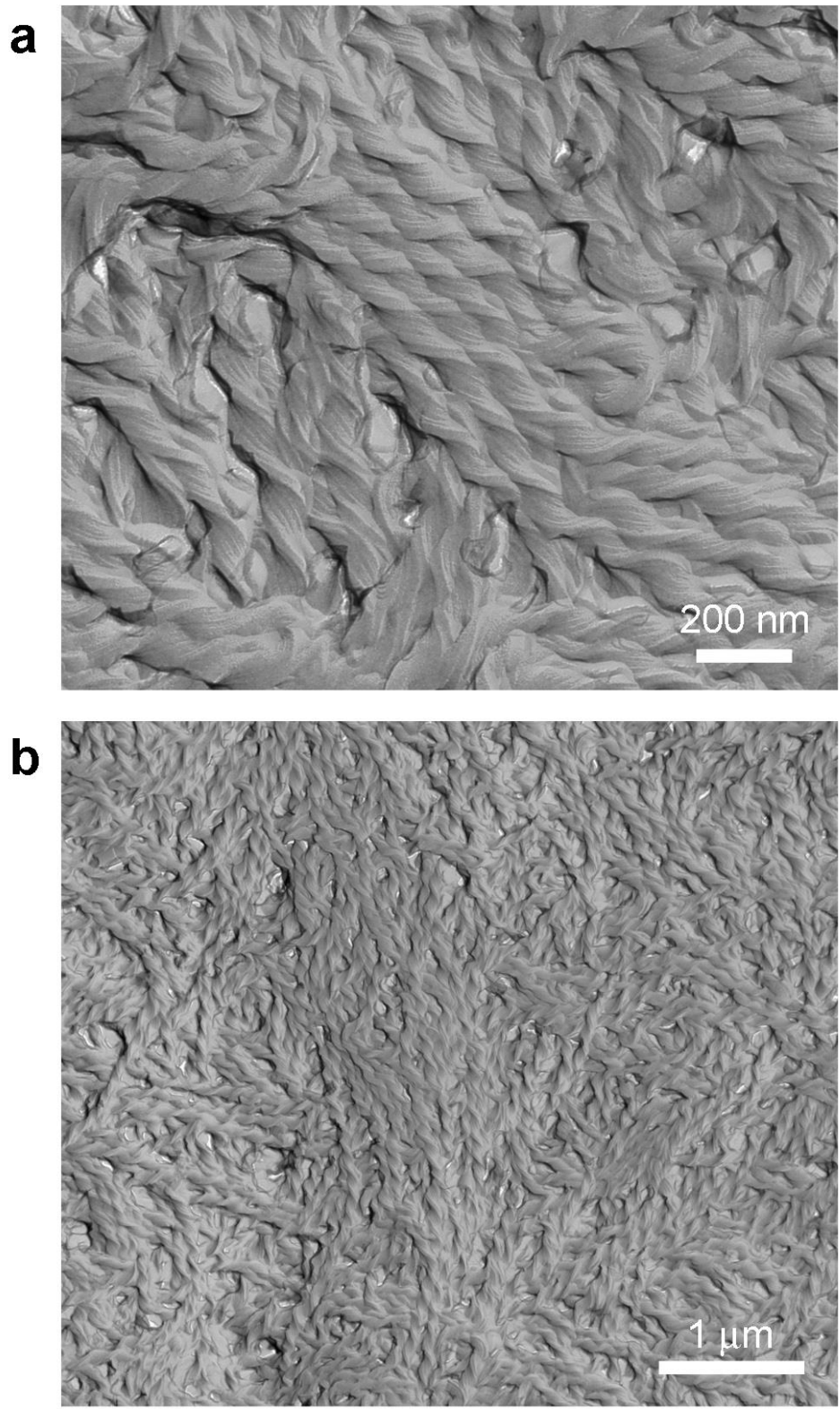

Figure S5. FFTEM images showing the HNF structure of 12OAz5AzO12 prepared by mixing with 60 wt% ZLI-2293. (a) 5 to 6 HNFs with coherent twist can be seen in the center of the image. (b) A large-scale view shows that the filaments can grow independently or collectively in the bulk and form a nanoporous structure.

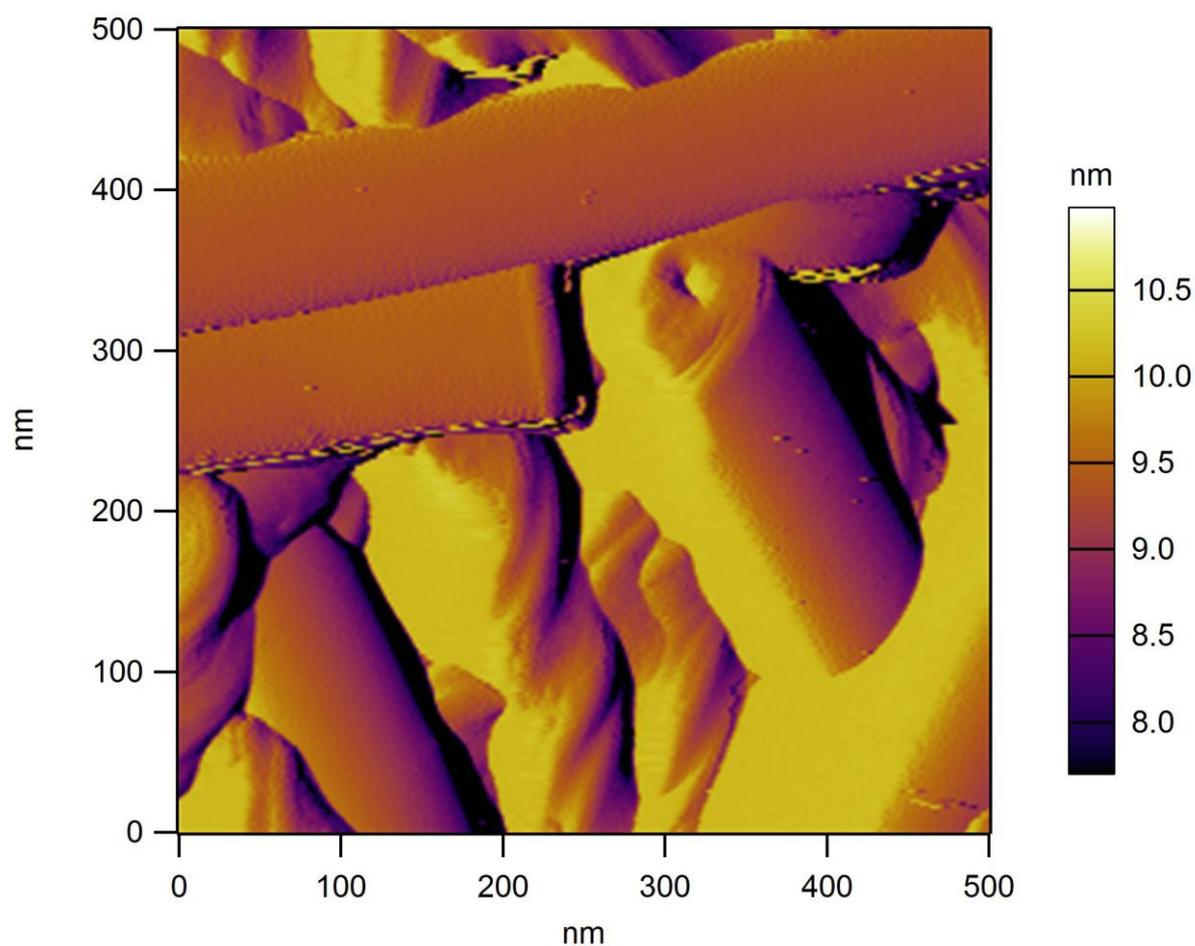

Figure S6. AFM image of the nanotube structure of 12OAz5AzO12. This image shows that nanotubes have essentially atomically smooth surfaces, while the HNFs at the bottom exhibit the smectic layers or smectic layer 'shelves' as curved lines. A nanotube with wrapped layers can also be seen when observed in an oblique direction.

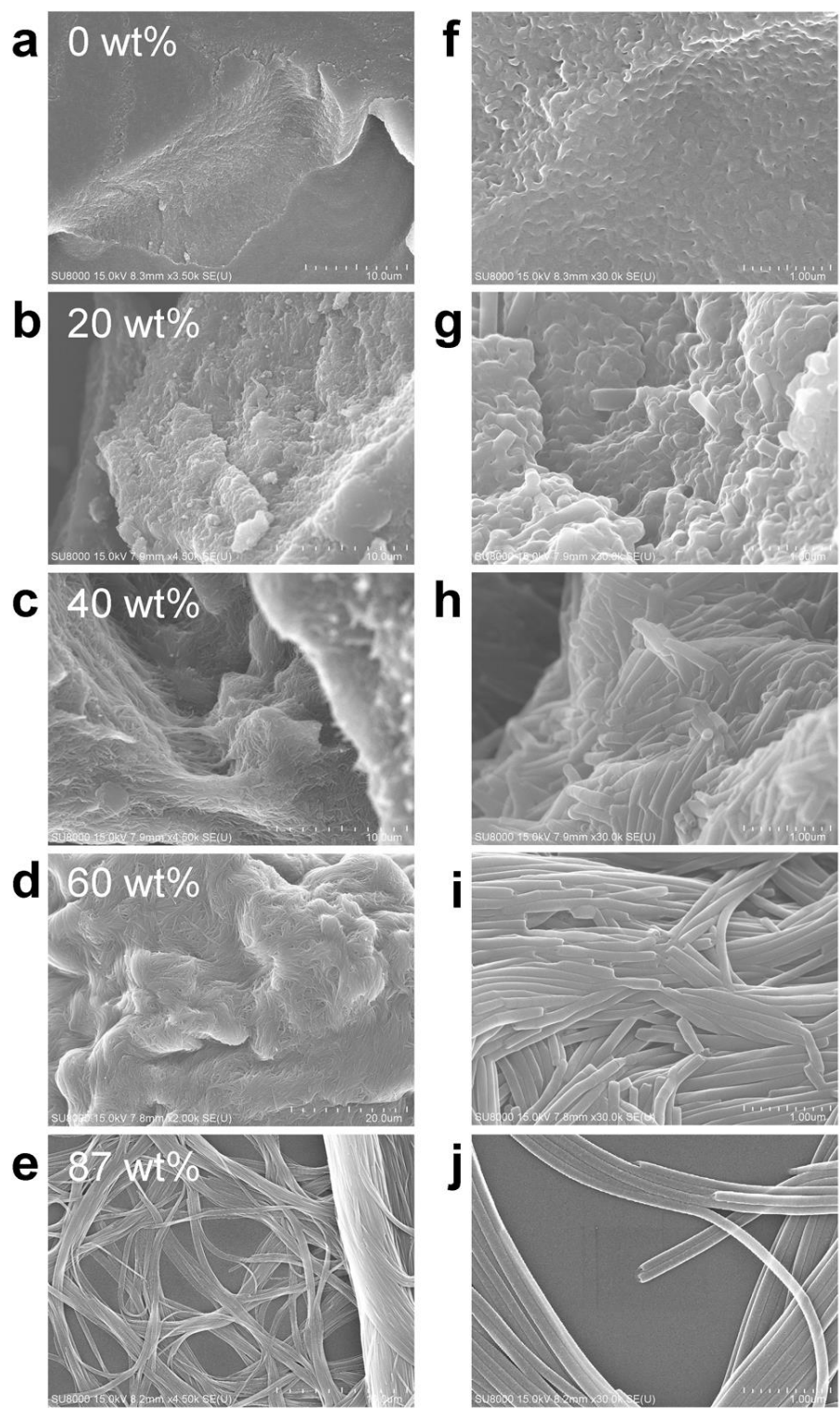

Figure S7. (a-e) A series of SEM images displaying the different nanostructures of the nanophase-separated 12OAz5AzO12 formed by mixing it with different amounts of n-dodecane (where 0 wt% refers to the neat 12OAz5AzO12), and (f-j) zoomed views of the image to its left. Note that the nanotube structure begins to emerge at lower n-dodecane concentrations (ca. 40 wt%), as compared with mixtures with ZLI-2293 (80 wt%, Fig. 3d). Scale bar 10 μm (except 20 μm for 60 wt %) in (a-e), and 1 μm in (f-j) images.

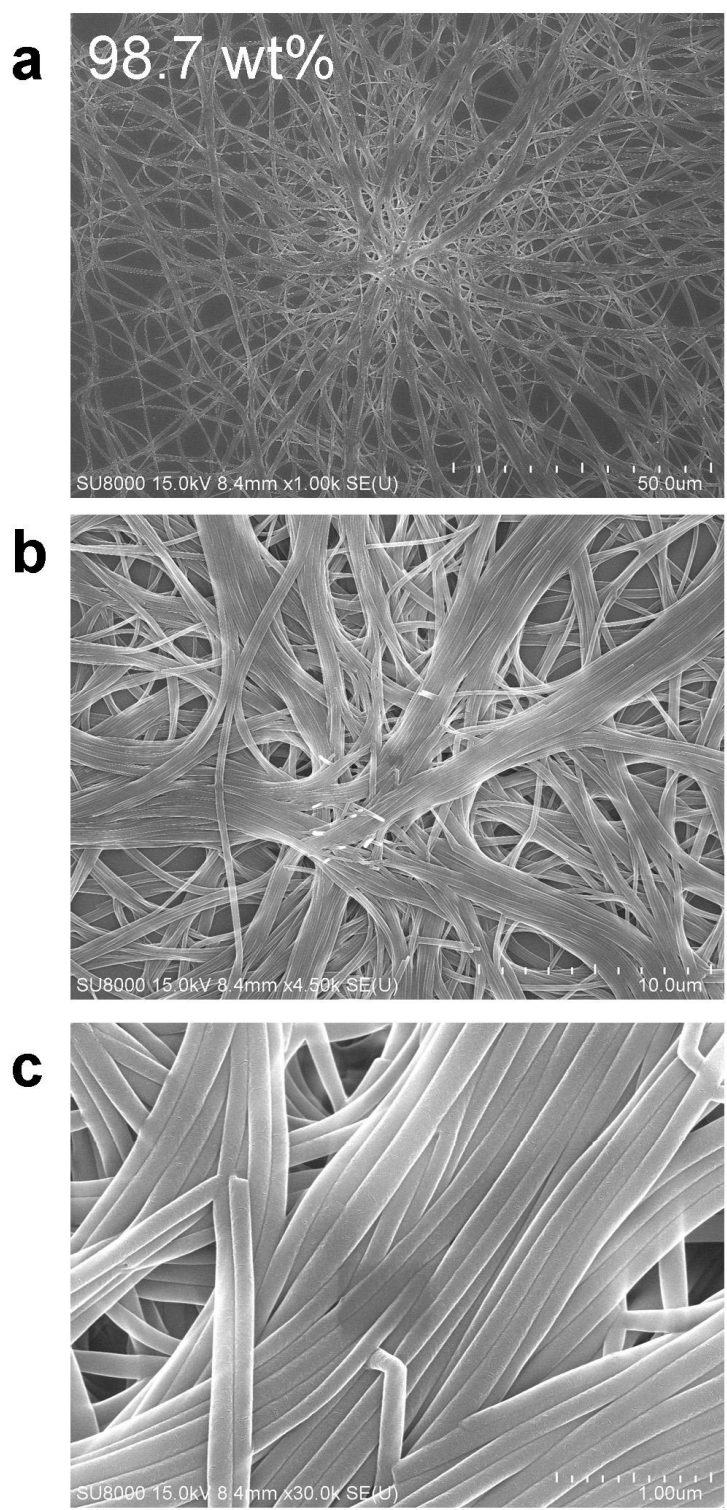

Figure S8. SEM images of the same region of nanotubes of 12OAz5AzO12 formed by mixing with 98.7 wt% n-dodecane with different magnifications. Scale bars are 50, 10, 1 μm in (a), (b), (c), respectively. The dark rectangle in the center of (c) is a burn caused by prolonged irradiation of the electron beam while adjusting the focus.

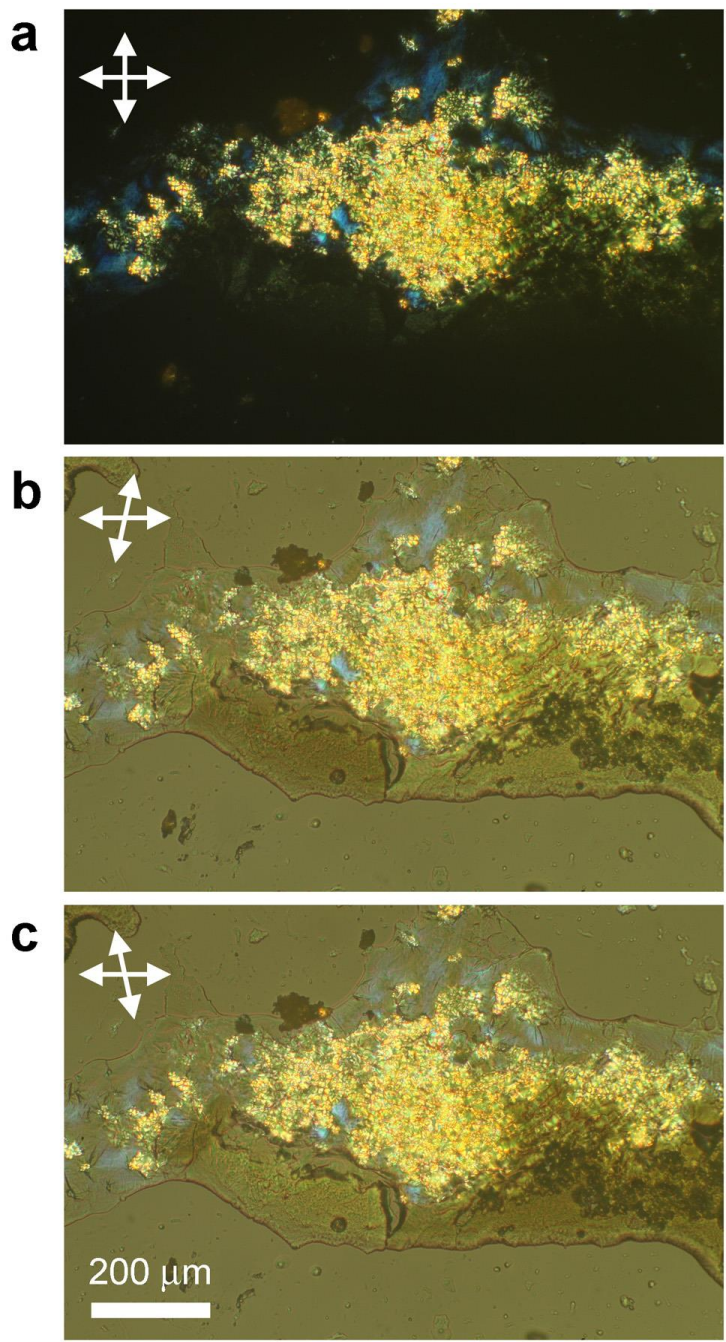

Figure S9. (a) POM textures reveal that when the composite SmX phase (which we believe to be in the DC phase in this case) is slowly heated up to the SmC$_A$ phase (yellow domains), it may re-crystallize and form blue domains. (b,c) Decrossing the polarizers reveals optical rotation in the crystal domains.

25